\documentstyle[pra,aps,preprint]{revtex}
\input{epsf}
\begin{document}
\title{A quantum-mechanical study of optical regenerators based on
nonlinear-loop mirrors}
\author{Giacomo M. D'Ariano\cite{dar} and Prem Kumar}
\address{Department of Electrical and Computer Engineering \\
Northwestern University, Evanston, IL 60208-3118, USA} 
\date{October 1, 1997; Revised: January 7, 1998}
\maketitle
\begin{abstract}
We present a quantum-mechanical analysis of a nonlinear interferometer
that achieves optical switching via cross-phase modulation resulting
from the Kerr effect. We show how it performs as a very precise
optical regenerator, highly improving the transmitted bit-error rate
in the presence of loss. 
\end{abstract}
\vfill
\centerline{\small To appear in Photonics Technology Letters}
\newpage
The increasingly large demand for information capacity in
telecommunication networks urges conversion of hybrid electro-optic
signal processing to all-optical processing, exploiting directly the
large bandwidth available in 
the optical domain. For long-distance communication, losses along 
the line represent the crucial limitation to the maximum transmission
span, and loss compensation through linear optical 
amplification unavoidably introduces noise of quantum-mechanical
origin\cite{Caves,Yuen}. However, when the data are digitally encoded,
instead of linear amplifiers one can effectively employ {\em digital
optical switches} or {\em optical regenerators}. Very
promising as optical switches are the nonlinear loop mirrors, which are 
nonlinear Sagnac interferometers made with cross-phase modulating Kerr
medium. Experimental studies at high power levels have shown that such
devices can achieve very effective switching, and optical regeneration
\cite{lucek93}
up to 40 Gbit/s has been practically demonstrated \cite{Pender96}. The
quantum characteristics of nonlinear Mach-Zehnder (or equivalently,
Sagnac) interferometers have been previously investigated~\cite{sanders92}.
In this Letter we address the performance limits for optical switching 
with such devices, and present numerical results for the
bit-error rate (BER) achievable in a lossy line with optical regenerators
distributed along the line.
\par A fiber-optic nonlinear Sagnac interferometer is depicted in
Fig.~\ref{f:loop} along with its equivalent Mach-Zehnder
interferometer. The input mode $\hat a$, assumed to be in a coherent
state $|\alpha\rangle$, is split into the two arms of the
interferometer by a 50/50 beam splitter. In one arm the pertaining
field mode ${\hat a}''$ undergoes a Kerr nonlinear phase shift
$\hat U=\exp[i\chi\hat {a''}^{\dag}\hat a'' \hat c^{\dag}\hat c]$ that
depends on the state of the other input mode $\hat c$ of a different
frequency and/or polarization. Here, by the
hooded letters $\hat a,\hat b,\hat c$ we are denoting the annihilation
operators of the respective field modes, by their daggered letters we
mean the respective creation operators, and $\chi$ denotes the overall
Kerr coupling, i.e., the third-order susceptibility of the medium 
integrated over its length along the direction of propagation. For
$\hat c$ in the vacuum state there is no nonlinear phase shift for 
$\hat a''$, and for an appropriate choice of phases in the two arms of
the interferometer the field modes exactly recombine at the second
50/50 beam splitter, so that the input coherent state $|\alpha\rangle$
emerges as the state for the output mode $\hat a'$. If  
$\hat c$ is non-vacuum, then a Kerr phase shift of magnitude $\pi$ would
make the state $|\alpha\rangle$ switch toward the output $\hat
b'$. Strictly speaking, a perfect $\pi$ phase shift is achieved only
with $\hat c$ in a number state $|n\rangle$, 
with $n\chi=\pi$. However, as we will see in the following, the
interferometer switches effectively even in the presence of 
Poissonian intensity fluctuations, i.e., when $\hat c$ is in a coherent
state, say $|\beta\rangle$, with $|\beta|^2\chi=\pi$. The Sagnac
interferometer in Fig.~\ref{f:loop}, when used as a repeater, is 
regarded as having the input at port $c$ and the output at port
$b'$. When the coherent state
$|\beta\rangle$ corresponding to bit {\em one} enters port $c$ the
interferometer approximately re-transmits the strong coherent state 
$|\alpha\rangle$ of the local-laser (LL) mode $\hat a$; otherwise, it 
just re-transmits the vacuum state corresponding to bit {\em zero}. 
\par The relation between the input state $\hat\rho_{{\rm in}}$ at $\hat c$
and the output state $\hat\rho_{{\rm out}}$ at $\hat b'$ of the Sagnac
repeater is easily derived; it is given by the following map:  
\begin{eqnarray}
&&\hat\rho_{{\rm out}}=\Gamma_{\chi,\alpha}(\hat\rho_{{\rm in}})\label{G}\\
&\equiv&\sum_{n=0}^{\infty}\Big|\alpha e^{{i\over2}
n\chi}\sin(n\chi/2)\Big\rangle\langle n|
\hat\rho_{{\rm in}}|n\rangle\Big\langle\alpha
e^{{i\over2}n\chi}\sin(n\chi/2)\Big|\nonumber\;,
\end{eqnarray}
which shows that the resulting output state is a mixture of coherent
states. When the input is in a high mean-intensity coherent state 
$\hat\rho_{{\rm in}}=|\beta\rangle\langle\beta|$ with
$|\beta|^2=\pi/\chi\gg1$, then the output state in Eq.~(\ref{G}) is
approximated by the dephased coherent state 
\begin{eqnarray}
\hat\rho_{{\rm out}}\simeq\int_{-\infty}^{+\infty}\!\!
\frac{d\varphi}{\sqrt{2\pi\Delta^2}}
e^{-\frac{\varphi^2}{2\Delta^2}}\Big|i\alpha
e^{i\varphi}\cos\varphi\Big\rangle\Big\langle
i\alpha e^{i\varphi}\cos\varphi\Big|,\!\!\!\label{cdephase}
\end{eqnarray}
having a small variance $\Delta^2=\pi^2/(4|\beta|^2)=\pi\chi/4\ll 1$.
The state in Eq.~(\ref{cdephase}) has a banana-shaped Wigner function,
corresponding to a quasi-Poissonian photon-number distribution with
$\langle\hat b^{\dag}\hat b\rangle\simeq |\alpha|^2\exp(-\Delta^2)$ 
and a Gaussian phase distribution with $\langle \Delta\phi^2\rangle=
|\alpha|^{-2}+\Delta^2$. 
\par We now consider an on-off communication scheme implemented on a
lossy line with transmitted average power $P=|\gamma|^2/2$, and the
{\em zero} and the {\em one} bit represented by the vacuum state
$|0\rangle$ and the coherent state $|\gamma\rangle$,
respectively. After a loss $1-\eta$ we insert a repeater with its Kerr
coupling tuned to the switching value $\chi=\pi/(\eta|\gamma|^2)$ and
the LL amplitude set to $\alpha=-i\gamma$, such that the original
amplitude $\gamma$ is 
approximately re-established at the output [for the $-i$ phase factor
see Eq.~(\ref{cdephase})]. After a further loss $1-\eta$ another repeater  
is inserted, and so forth, for an overall $N$ number of steps (see
Figs.~\ref{f:x-loss} and~\ref{f:x-line}). Since
the only effect of loss on coherent states is just an amplitude 
rescaling by the factor $\sqrt{\eta}$, and because the output
state from the Sagnac interferometer is a mixture of coherent states,
it turns out that the overall state transformation for a loss $1-\eta$ 
preceded by a repeater depends only on the signal level $\beta$ 
at the repeater input, independently of $\eta$, as long as the LL is
set at the loss-compensating amplitude value $-i\eta^{-1/2}\beta\equiv
-i\gamma$ and the Kerr coupling is tuned to the switching value
$\chi=\pi/|\beta|^2$. Hence the overall input-output map for the
repeater-loss sequence is given by $\Gamma_{\pi/|\beta|^2,-i\beta}$.
The fact that this map depends only on the input signal level $\beta$
does not mean that one can recover a given signal after any
arbitrarily-high loss. If the input signal level is reduced too much,
i.e., if $\beta\equiv\eta^{1/2}\gamma\to 0$ for a fixed peak 
amplitude $\gamma$, then the dephasing effect at the repeater would
increasingly degrade the carrier coherence, leading to enhanced fluctuations
$\Delta^2=\pi^2/(4\eta|\gamma|^2)$ for $\eta\to0$. 
\par We now evaluate the transmitted BER for on-off keying
and direct photodetection at the end of $N$ optical regenerators
distributed along the lossy line. The input is in the coherent state
$|\beta\rangle$ and all repeaters are set at their optimal working
point with the Kerr coupling $\chi=\pi/|\beta|^2$, and with the $n$th
repeater having a LL intensity given by $|\alpha_n|^2=|\beta|^2/\eta_n$, 
where $1-\eta_n$ is the loss between the $n$th and the $(n+1)$th repeater.
In this scheme, since there is no spontaneous emission from the repeater,
the detection threshold should be set at one photon, and the BER is just
the vacuum component of the 
output state after iterating the map $\Gamma_{\pi/|\beta|^2,-i\beta}$
in Eq.~(\ref{G}) $N$ times on the input state $|\beta\rangle$.
A numerical calculation shows that after a sudden dephasing at the
first step, the state remains quite stable with  
a slow dephasing in the following steps. A log-log plot of the BER $B$
versus $N$ is reported in Fig.~\ref{f:ber} for different magnitudes of
the input amplitude $\beta$. The BER increases fast for the first
ten-twenty steps. Then, there is a transition to a very stable linear
regime $B=C(\beta)N$, with $C(\beta)=10^{-0.44\pm
0.02-(0.0356\pm 0.0001)|\beta|^2}$. So, for example, after $N=10^2$ 
optical repeaters and for input amplitude $\beta=20$ the BER is 
$B=2\times 10^{-13}$. This result should be compared with the BER
achieved on a lossy line with ideal photon-number amplifiers
\cite{yuen2}, where for a gain $g=\eta^{-1}=2$ the BER is 
$B=3.4\times 10^{-7}$.
The performance is much worse when using linear phase-insensitive
amplifiers due to the spontaneous-emission noise. 
\par In conclusion we analyzed a nonlinear Sagnac interferometer with
cross-phase modulation, which is used as a repeater for on-off modulation 
and direct detection in a lossy line with $N$ distributed repeaters. We
showed that with all repeaters set at their optimal working 
point, the BER increases linearly with $N$ for large $N$, and the
proportionality constant exponentially decreases with the input 
signal intensity, resulting in almost error-free communication even at
very low power levels. Finally, our monochromatic analysis remains valid
even for short pulses as long as dispersion effects are negligible
and the cross-Kerr susceptibility can be considered approximately
constant over the pulses' frequency bandwidth. 
\par This research is sponsored by Defence Advanced Research Projects
Agency and Rome Laboratory, Air Force Materiel Command, USAF, under
agreement number F30602-97-1-0240. The U.S.\ Government is authorized
to reproduce and distribute reprints for Governmental
purposes notwithstanding the copyright annotation thereon.
The views and conclusions contained herein are those of the authors
and should not be interpreted as necessarily the official policies or
endorsements, either expressed or implied, of the DARPA, Rome
Laboratory, or the U.S.\ Government.

\begin{figure}
\caption{Schematic of a nonlinear fiber-optic Sagnac interferometer (a),
and its Mach-Zehnder equivalent (b). WDC, wave-length dependent
coupler.}
\label{f:loop}
\end{figure}
\begin{figure}
\caption{Schematic representation of the nonlinear Sagnac
interferometer used as a regenerator along a transmission line. 
Considering only the relevant ports, the regenerator is a three-port 
device. The optimal working point for the local laser is shown, which 
depends on the loss following the regenerator.}
\label{f:x-loss}
\end{figure}
\begin{figure}
\caption{Schematic of a lossy line with distributed optical regenerators.
\hspace*{1.25in}}
\label{f:x-line}
\end{figure}
\begin{figure}
\caption{Bit-error rate $B$ for a lossy line with $N$ optical repeaters 
versus the number $N$ of repeaters (see text). Different curves refer
to different input field amplitudes $\beta$, ranging in unit steps from the
bottom curve that corresponding to $\beta=17$. The slowly increasing
regime for large $N$ is linear, and its form is reported in the figure
inset.}
\label{f:ber}
\end{figure}
\newpage
\begin{figure}[p]
\centerline{\epsfxsize=4.5in \epsfbox{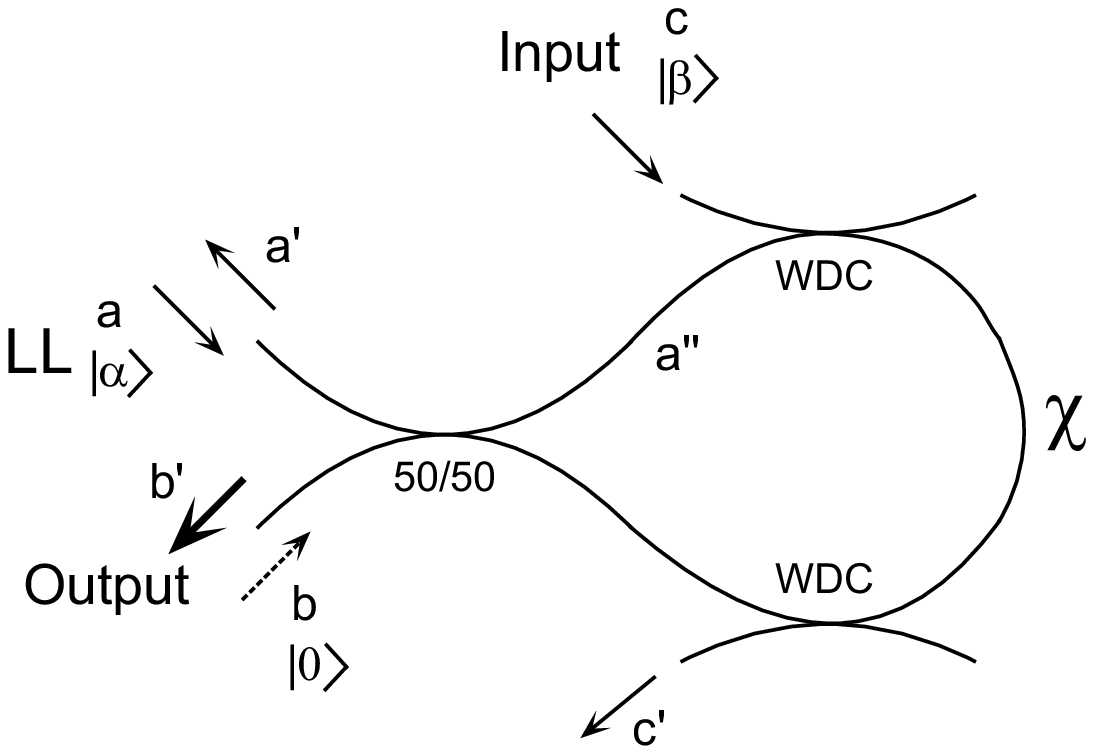}}
\vspace*{0.25in}
\centerline{\bf (a)}
\vspace*{0.5in}
\centerline{\epsfxsize=4.5in \epsfbox{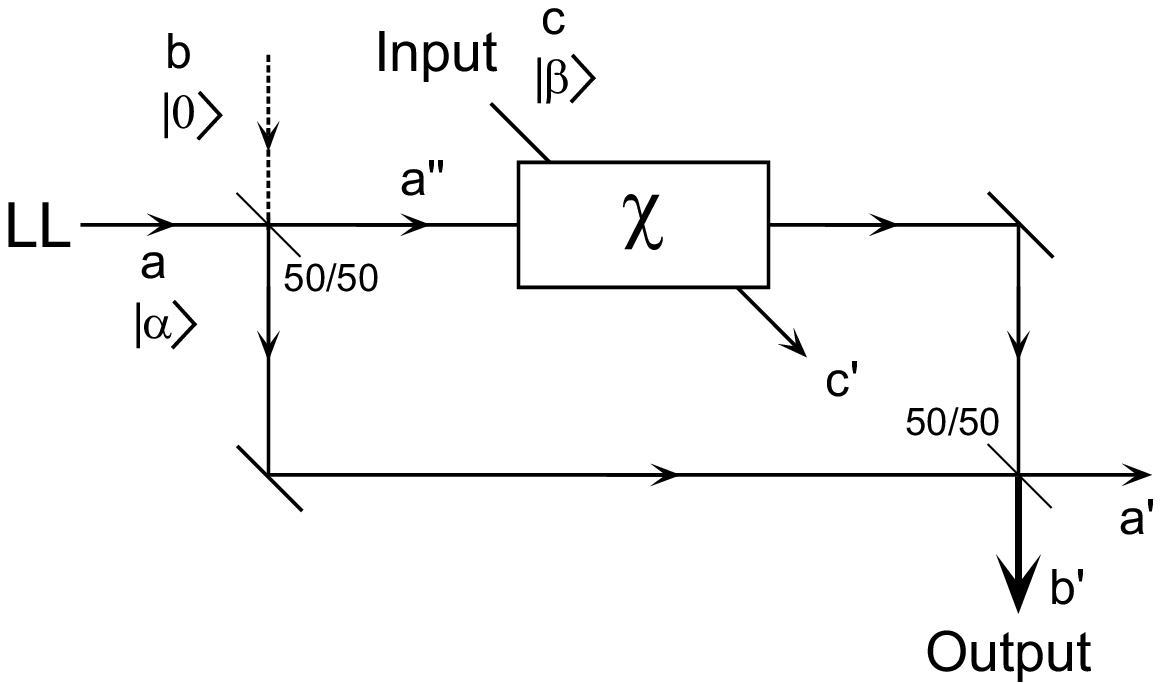}}
\centerline{\bf (b)}
\vfill
\rightline{{\tt Fig.~1} ~ G. M. D'Ariano and P. Kumar}
\rightline{{\em A quantum-mechanical study of optical regenerators ...}}
\end{figure}    
\newpage
\begin{figure}[p]
\vspace*{2in}
\centerline{\epsfxsize=5in \epsfbox{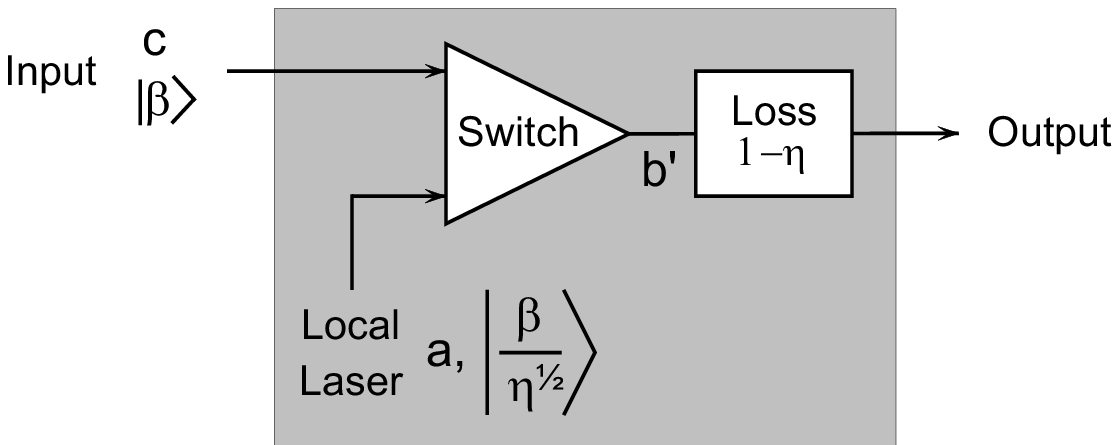}}
\vfill
\rightline{{\tt Fig.~2} ~ G. M. D'Ariano and P. Kumar}
\rightline{{\em A quantum-mechanical study of optical regenerators ...}}
\end{figure}     
\newpage
\begin{figure}[p]
\vspace*{2in}
\centerline{\epsfxsize=6.5in \epsfbox{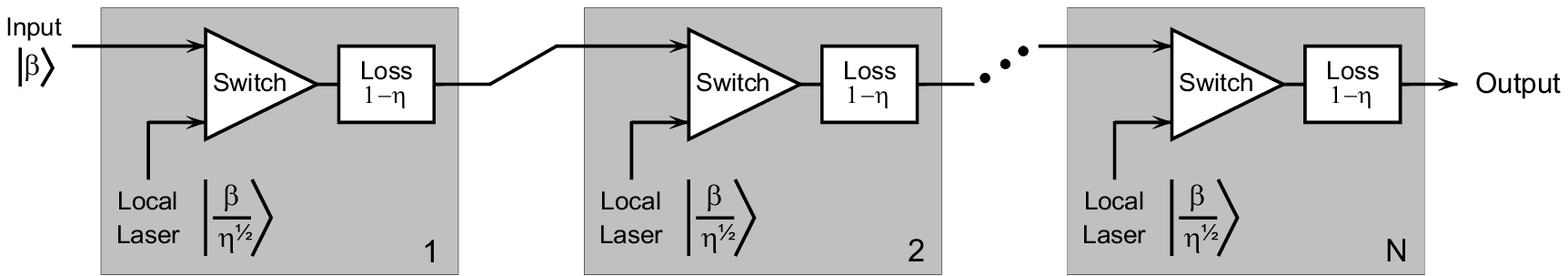}}
\vfill
\rightline{{\tt Fig.~3} ~ G. M. D'Ariano and P. Kumar}
\rightline{{\em A quantum-mechanical study of optical regenerators ...}}
\end{figure}     
\newpage
\begin{figure}[p]
\vspace*{1.5in}
\centerline{\epsfxsize=5in \epsfbox{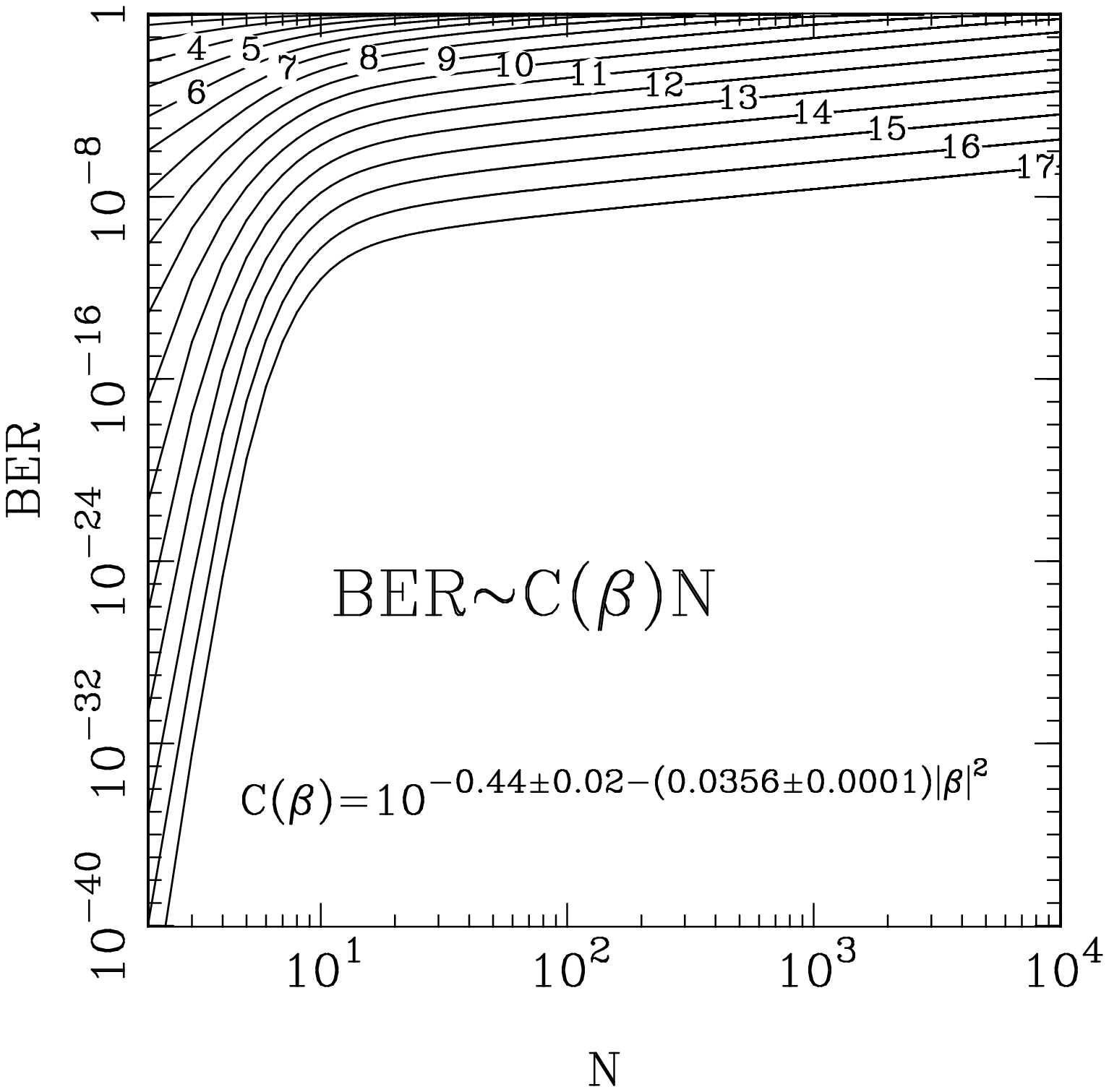}}
\vfill
\rightline{{\tt Fig.~4} ~ G. M. D'Ariano and P. Kumar}
\rightline{{\em A quantum-mechanical study of optical regenerators ...}}
\end{figure}     

\begin{references}
\bibitem[*]{dar} Permanent address: Dipartimento di Fisica 
``A. Volta,'' via Bassi 6, I 27100 Pavia, Italy.
\bibitem{Caves} C. M. Caves, ``Quantum limits on noise in linear 
amplifiers,'' {\em Phys.\ Rev.\ D}, vol.\ 26, pp.\ 1817--1839, 1982.
\bibitem{Yuen} H. P. Yuen, ``Design of transparent optical networks 
by using novel quantum amplifiers and sources,''
{\em Opt.\ Lett.}, vol.\ 12, pp.\ 789--791, 1987.
\bibitem{lucek93} J. K. Lucek and K. Smith, ``All-optical signal
regenerator,'' {\em Opt.\ Lett.}, vol.\ 18, pp.\ 1226--1228, 1993.
\bibitem{Pender96} W. A. Pender, T. Widdowson, and A. D. Ellis,
``Error-Free Operation of a 40 Gbit/s All-Optical Regenerator,''
{\em Electron.\ Lett.}, vol.\ 32, pp.\ 567--568, 1996.
\bibitem{sanders92} B. C. Sanders and G. J. Milburn,
``Quantum limits to all-optical switching in the nonlinear 
Mach-Zehnder interferometer,''
{\em J. Opt.\ Soc.\ Am.\ B}, vol.\ 9, pp.\ 915--920, 1992.
\bibitem{yuen2} H. P. Yuen, {\em Phys.\ Lett.}, vol.\ A113,
pp.\ 405--407, 1986.
\end{references}
\end{document}